\documentclass[prb,11pt,twocolumn,superscriptaddress,singlelinecheck=false]{revtex4}

\usepackage[colorlinks]{hyperref}
\usepackage{natbib}
\usepackage{amsmath}    
\usepackage{graphicx}   
\usepackage{verbatim}   
\usepackage{color}      
\usepackage{subfig}  
\usepackage{hyperref}   
\begin{document}

\title{Absolute Position Total Internal Reflection Microscopy with an Optical Tweezer }
\author{Lulu Liu}
\email[Email: ]{lululiu@fas.harvard.edu}
\author{Alexander Woolf}
\affiliation{School of Engineering and Applied Sciences, Harvard University}
\author{Alejandro W. Rodriguez}
\affiliation{Department of Electrical Engineering, Princeton University}
\author{Federico Capasso}
\email[Email: ]{capasso@seas.harvard.edu}
\affiliation{School of Engineering and Applied Sciences, Harvard University}

\begin{abstract}
A non-invasive, in-situ calibration method for Total Internal Reflection Microscopy (TIRM) based on optical tweezing is presented which greatly expands the capabilities of this technique.  We show that by making only simple modifications to the basic TIRM sensing setup and procedure, a probe particle's absolute position relative to a dielectric interface may be known with better than 10 nm precision out to a distance greater than 1 $\mu$m from the surface. This represents an approximate 10x improvement in error and 3x improvement in measurement range over conventional TIRM methods.  The technique's advantage is in the direct measurement of the probe particle's scattering intensity vs. height profile in-situ, rather than relying on calculations or inexact system analogs for calibration.  To demonstrate the improved versatility of the TIRM method in terms of tunability, precision, and range, we show our results for the hindered near-wall diffusion coefficient for a spherical dielectric particle.  
\end{abstract}

\maketitle

\section{Introduction}

Total Internal Reflection Microscopy (TIRM) is a near-field optical sensing technique which allows for the detection of surface forces down to the sub piconewton regime with nanometer resolution\cite{p99}.  In TIRM, a particle is tracked by the intensity of light it scatters from the evanescent field of a totally internally reflected beam.  Since the intensity of an evanescent field drops off exponentially with distance from the interface $z$, that is,
\begin{align}
\label{exponentialIz}
I_{field} (z) = I_0 e^{-\beta z}
\end{align}
with $\beta^{-1}$ typically 100 nm or less, the technique is among the most sensitive for monitoring particle motion near a surface. By measuring the bead's height probability density function in thermal equilibrium and inverting the Boltzmann distribution, the potential energy profile can easily be obtained\cite{p99}.  As such, TIRM has been utilized with good success to measure electrostatic double layer, van der Waals, optical, and critical Casimir forces acting on micron-sized particles \cite{p90,b99,f93,w92,he08}.  Since TIRM is also non-invasive and non-destructive, enabling tracking and sensing with very low laser powers ($\sim$1 mW), it is also a popular choice among biologists\cite{a81,l95,a01}.  However, in measurements over a large range of distances from the surface, requiring accurate knowledge of absolute positions, or involving highly reflective materials such as metals, the potential of the technique has not been fully realized \cite{c01,b05,vo06}.  In traditional TIRM experiments, measurements proceed uncalibrated.  The intensity of the light scattered by the probe particle as a function of height is essentially unknown and assumed to follow the exponential profile in Eq. \ref{exponentialIz} with the decay length ($\beta^{-1}$) calculated from the geometry of the setup.

Past efforts at calibrating the intensity-height profile have gone in two main directions.  The first replaces the dynamical experimental system with a carefully fabricated, static, calibration standard which mimics experimental parameters but has known height values \cite{pr93, ma06,ge09,mc05}.  Other methods exploit calibrations based on known hydrodynamic interactions\cite{vo09} or diffusion dynamics\cite{ha04}.  The main disadvantage of these existing approaches is their large experimental error.  Experimental parameters may be estimated and reproduced only to a certain extent (see Section \ref{sec:betas}).  Currently, the aim to keep calibration uncertainty within bounds has led to the exclusion of certain novel or difficult-to-target configurations, such as metal surfaces or larger evanescent field penetration depths ($\beta^{-1} > $ 150 nm).   In most cases, absolute positional values are forfeited altogether in favor of relative displacement as existing calibration methods are labor-intensive and imprecise.

We propose an optical tweezer\cite{as86}-based calibration which directly measures the intensity-height profile for a given scatterer and TIRM configuration and can be performed quickly and in-situ.  The optical tweezer, or single-beam gradient optical trap, holds the scatterer in three dimensions at a fixed position relative to the beam focus \cite{as92}.  By shifting the focus, the particle can be made to approach the surface in precisely measured steps. The scattering intensity is monitored until further steps produce no change in the signal, indicating that the bead has reached the surface.  

In section \ref{sec:exp}, we discuss the experiment in detail. Section \ref{sec:setup} describes the setup.  A few straight-forward changes made to the usual TIRM setup enables our in-situ calibration.  First, the collection objective is replaced by a high NA water-immersion objective capable of 3D optical tweezing of a micron-sized dielectric particle.  Second, the micrometer vertical stage on which the objective is mounted is fitted with closed-loop piezo controls for finely calibrated focus adjustment.  And finally, the top surface of the glass sample slide is coated with a quarter-wavelength thick layer of evaporated glass to cancel the optical tweezer reflected beam.  By opening and closing an iris located in the back focal plane of the objective, we adjust trap laser power and NA and can quickly switch between 3D optical tweezing (calibration) and 2D optical trapping (measurement) modes of operation.  

In Section \ref{sec:data}, we discuss the data-taking procedure and detailed considerations that make this calibration possible.  Since the correspondence of piezo step-size to particle displacement is an important prerequisite in our experiment, we take steps to measure and eliminate distortions in the particle's trapping potential energy profile due to reflections or the presence of a thick electrical double layer.  Data is presented which demonstrates the necessity and effectiveness of an anti-reflection coating as well as the addition of salt into the suspending fluid.

In section \ref{sec:results}, we demonstrate the effectiveness of the proposed calibration method for absolute position TIRM with two main results.  The first experiment measures intensity-height calibration curves for several evanescent field penetration depths ranging from 90 nm to 270 nm taken with the same probe particle.  The error on $\beta$ determined in this way does not diverge as the angle of incidence approaches the critical angle as with traditional methods, but rather remains below 1\% in all fits.  A key and yet unexplored feature of TIRM sensing techniques is the ability to freely tune (even in real time) the length-scale parameter $\beta$ to optimize for precision or range. Our demonstrated ability to measure, accurately and in-situ, the $\beta$ parameter should therefore greatly improve the versatility and applicability of the TIRM technique.  

To prove this extended versatility, along with the calibration's accuracy, we also perform experiments in which we fix the angle of incidence such that the penetration depth is about 300 nm and measure the hindered perpendicular diffusion coefficient of a 3 $\mu$m glass bead in steps of 50 nm out to a distance of about 1.2 $\mu$m.  Our results agree with hydrodynamic theory throughout, and is, to our knowledge, the longest range and most accurate measurement of hindered near-wall diffusion using TIRM.

\section{Experiment}
\label{sec:exp}
\subsection{Setup}
\label{sec:setup}

\begin{figure*}[t!]%
\centering
\subfloat[][]{\includegraphics[width=2.6in]{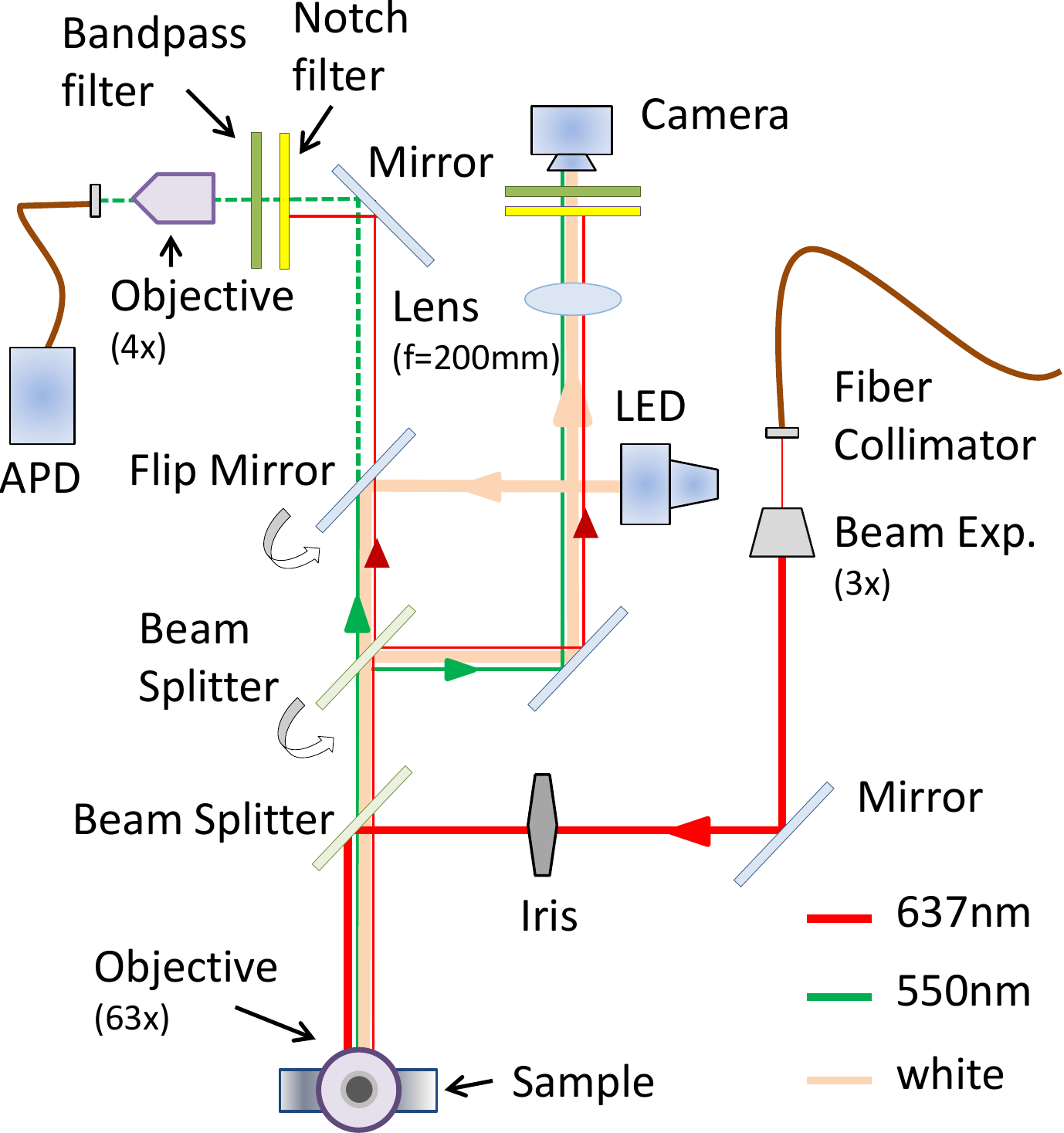}}%
\qquad
\subfloat[][]{\includegraphics[width=3.3in]{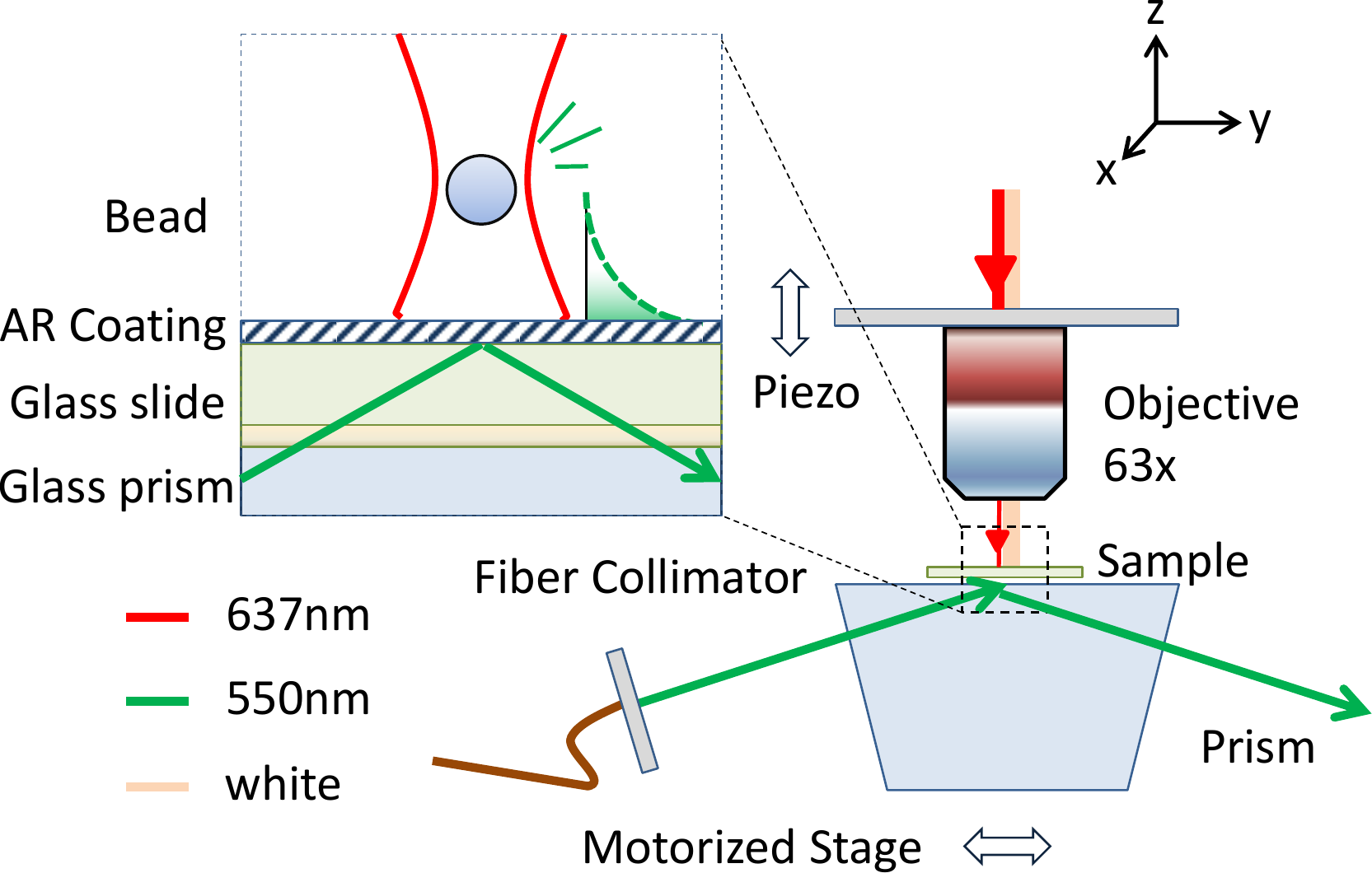}}%
\captionsetup{justification=raggedright,singlelinecheck=false}
\caption{ \footnotesize a) Top-down view of the total internal reflection microscopy (TIRM) setup with the trap, scattered and illumination paths denoted with red, green, and yellow lines respectively.  A 45$^\circ$ mirror above the 63x objective used to direct the beam downward onto the sample surface is not shown.  b) Side-view of the TIRM setup.  Inset: zoom in of a measured bead with the optical trap (red) and evanescent field profile (green) superimposed.}%
\label{fig:setup}%
\end{figure*}

In our experiment we use plain glass spheres with a density of 2.0 g/cm$^3$ and a nominal diameter of 3 $\mu$m (Corpuscular, Inc C-SIOs-3.0).  The beads come suspended in deionized water and are diluted by a factor of 100 and allowed to equilibrate at room temperature before being loaded into the sample chamber.  The chambers have thicknesses of 15-20 $\mu$m for use with a short working-distance objective and are fabricated by pressing two strips of sealing film (Solaronix Meltonix 1170-25) between a number 1.7 coverslip and a glass slide and heating to 120C for 3-4 hours in an oven.  The chambers are open on two sides so that the samples may be loaded by capillary action of the fluid. 

The anti-reflection (AR) coating, when it is applied, is deposited onto the top surface of the glass slide via plasma enhanced chemical vapor deposition (PECVD).  It consists of a quarter-wavelength thick layer of evaporated SiO$_2$, with index 1.45, which closely approximates the geometric mean of n=1.42 for glass (n=1.52) and water (n=1.33).

Figure \ref{fig:setup} shows the optical set up, a top-down and a side-view are provided.  Three light sources illuminate our sample.  The collimated output from a 637 nm diode laser (Thorlabs LP637-SF70) serves as our single beam optical trap.  It is expanded then focused through an objective onto the sample surface. The evanescent wave used for positional sensing is generated by total internal reflection of a 550 nm laser source (NKT Photonics SuperK) at the lower water-glass boundary of the sample chamber.  A polarizer and a half-wave plate is used to control the polarization of the 550 nm source.  Lastly, for visualization and positioning, a white LED provides bright-field illumination of the sample surface over our 50 $\mu$m field of view.

During the experiment, our sample slides rest on a 60 degree prism made of BK7 glass with index-matching fluid in between. The prism is mounted on a two-axis motorized translation stage (Thorlabs LNR50S) which controls lateral positioning of the sample with micron-precision.  A vertical translation stage (Thorlabs MAX301) with closed loop piezo electronics positions the objective with better than 5 nm precision.  The vertical travel range is 4 mm manually and 20 $\mu$m by piezo.

We use a water immersion objective (Leica 63x NA 1.25) in order to avoid the detrimental effects of spherical aberration introduced by the immersion oil and aqueous suspension fluid index mismatch \cite{ro02}.  This also ensures that a vertical step in the piezo motor corresponds to an equal magnitude displacement of the optical trap focus in our sample fluid.  

The same objective collects the photons scattered by a single glass bead from the evanescent sensing field. The scattered light, after traversing bandpass filters, is focused by a 4x objective onto a multimode optical fiber and fed into a photon counter (Micro Photon Devices PDM APD).  The photon counting interval can be set with a function generator in combination with a pulse counter (National Instrument PCI-6602) and in this experiment ranges from 200 $\mu$s to 1 ms.  The diameter of the multimode fiber core is 100 $\mu$m which sets the size of the collection area on the sample surface, measured with a stuck bead, to be about 6 $\mu$m in diameter.  Outside this area, the extinction is nearly complete, with counts at background levels.  Alignment is performed prior to each experiment to center the optical trap in the collection region.

\subsection{Data}
\label{sec:data}
\begin{figure*}[t!]%
\centering
\subfloat[][]{\includegraphics[width=3in]{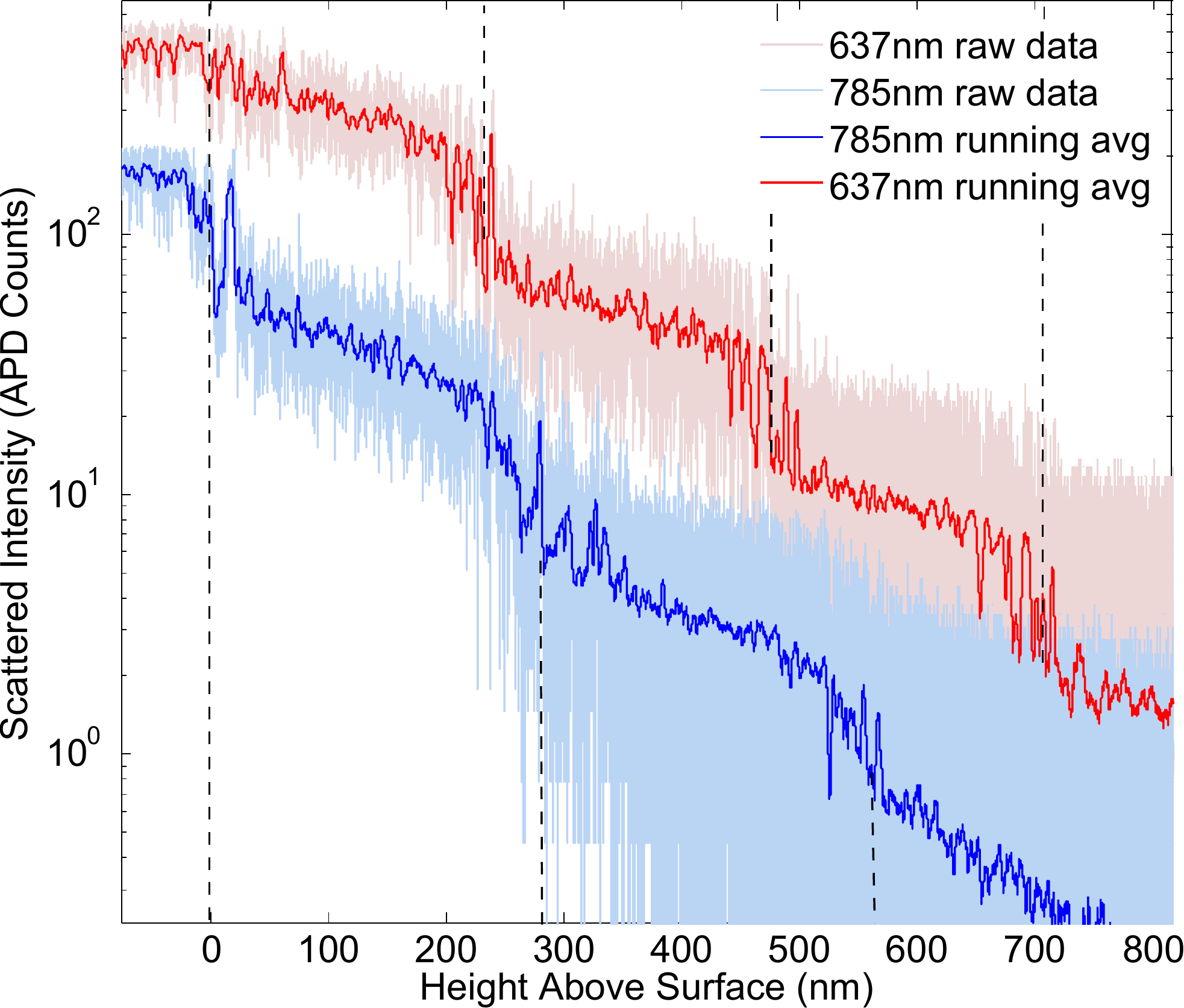}}%
\qquad
\subfloat[][]{\includegraphics[width=3in]{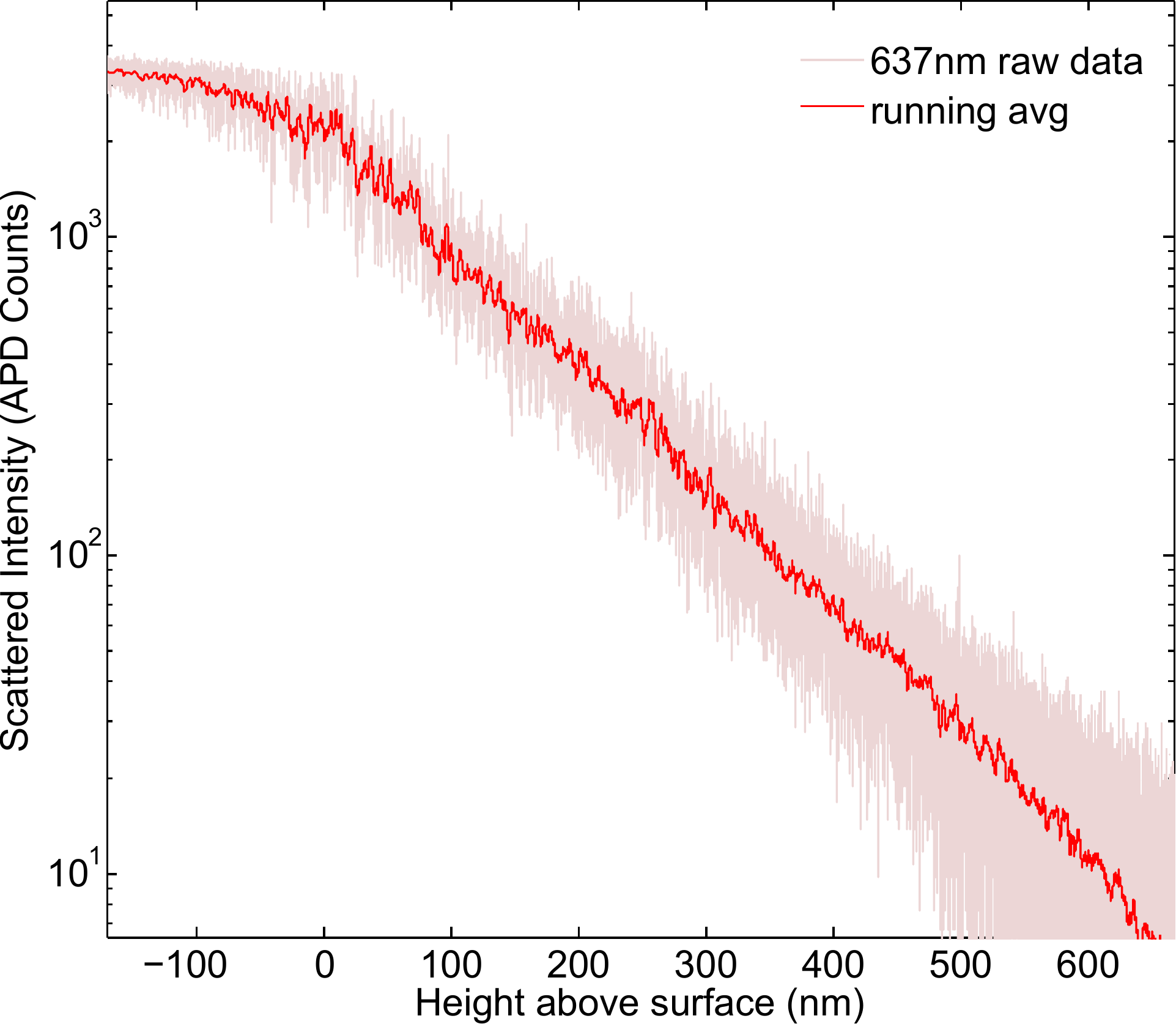}}
\captionsetup{justification=raggedright,singlelinecheck=false}
\caption{ \footnotesize  Plots of measured scattered light intensity versus height for a 3 $\mu$m bead above the surface of a glass microscope slide.  Red: using a 637 nm wavelength optical tweezer.  Blue: using a 785 nm wavelength optical tweezer.  (a) No AR coating: the step-like behavior is due to formation of a standing-wave component in the optical trap due to reflection from the glass-water interface.  The plots are shifted vertically for clarity.  (b) AR coated: the reflection is eliminated and the bead is able to step smoothly towards the surface.
}%
\label{fig:ar}%
\end{figure*}

We observe that the single-beam gradient optical trap \cite{as86} was sufficient to lift a 3 $\mu$m diameter glass bead in water at powers as low as 20 mW behind the objective.  Since in this experiment gravitational and radiation pressure forces act in the same direction, the bead, as expected, is trapped axially slightly below the focus of the beam \cite{as92}. The 550 nm laser source, used to generate the evanescent field for sensing near the glass-water interface, is kept at relatively low power (less than 1 mW at a spot size of 300 $\mu$m) to prevent perturbation of the optical field and optical forces acting on the bead. Background counts, taken with the bead held far (at least 2 $\mu$m) from the glass-water interface, are mostly due to scattering of the evanescent field by impurities on the glass slide, and contribute approximately 15 to 20 photon counts per millisecond, resulting in a signal-to-noise ratio typically better than 100 for our experiments.

The particle absolute position as a function of scattering intensity is determined as follows: the background measurement is taken, then the bead is lowered in precisely measured steps towards the surface and the scattered intensity is monitored until contact of the bead with the surface is made. Two abrupt changes in the collected signal are observed at the bead-surface contact point (Figure \ref{fig:ar}). First, the intensity becomes constant with further piezo steps, and second, the variance in measured counts decreases and approaches Poisson statistics. This decrease in variance is due to the fact that the bead is no longer diffusing since it is in contact with the surface. We use these features to find the location of the surface and therefore the absolute position of the bead. Since the ability of the bead to approach the surface is limited by the presence of a strong electrical double layer force \cite{p90,d91}, NaCl is introduced into the sample fluid at a minimum concentration of 0.01M in a typical calibration scan.  The effects of salt concentration on the measured intensity profile will be further discussed in Section \ref{sec:salt}. 

\subsubsection{AR Coating}
\label{sec:ar}

Optical tweezing near a dielectric boundary suffers from a known complication due to interference caused by a weakly-reflected backward-traveling wave \cite{c01,jo01}.  Therefore, even as the focus of the optical tweezer shifts, the interference fringes remain fixed relative to the location of the surface, resulting in step-like behavior of the trapped particle at regular intervals \cite{ja03}.  The effect of the standing wave modulation of our optical trap, despite the relatively low index contrast and reflectivity ($\sim$0.4\%) at the water-glass interface, can be observed in our data (Figure \ref{fig:ar}(a)). We further confirm the effect is optical by changing the wavelength of the trapping laser from 637nm to 785nm. The spacing of the fringes is measured to increase 21$\pm$2\% which agrees with the theoretical prediction of 23\%.

In order to overcome this complication we utilize the previously described AR coating made of a quarter-wave layer (110 nm thickness) of SiO$_2$, which reduces by more than a factor of 10 the standing-wave modulation of the optical tweezer while still maintaining the glass surface's chemical and electrical properties.

The intensity vs. height profile measured for a 3 $\mu$m glass sphere above an AR-coated glass surface is shown in Figure \ref{fig:ar}(b).  Generalization of this AR coating to different materials including highly reflective and metallic surface layers will be investigated in future work as it would allow for precise detection of novel optical forces, such as surface plasmon and repulsive casimir forces, with TIRM.

\subsubsection{Salt Concentration}
\label{sec:salt}

\begin{figure}
\captionsetup{justification=raggedright,singlelinecheck=false}
\includegraphics[width=3in]{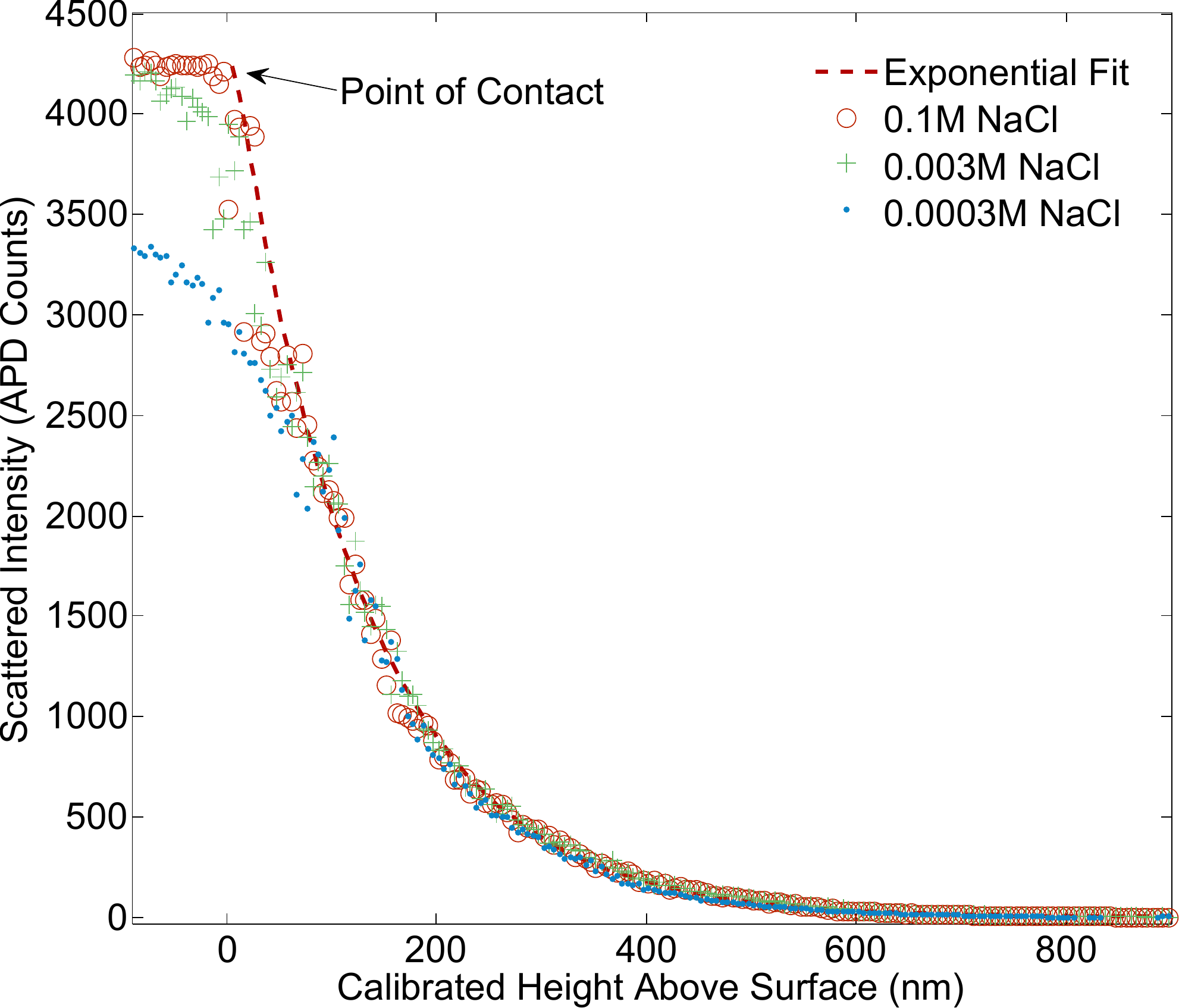}
\caption{ \footnotesize Intensity-height curves obtained for the a 3um diameter bead in water for various NaCl concentrations (corresponding Debye lengths: 1 nm, 5 nm, 20 nm).  The blue curve shows a condition of insufficient screening: despite the optical tweezer focus moving beneath the surface of the glass, the negatively charged bead is repelled electrostatically at close separations and unable to actually approach the negatively charged surface.  Green and red curves: additional salt brings calibration curves closer to the ideal exponential behavior. }%
\label{fig:salt}%
\end{figure}

Plain glass beads in DI water above a glass substrate exhibit strong electrical double layer repulsion since both surfaces acquire a negative charge upon contact with the fluid.  This electrostatic repulsion is partially screened by a mobile layer of positively charged counter-ions whose thickness (characterized by the Debye length) depends upon the concentration of dissolved electrolytes and can be hundreds of nanometers in extent in low molarity aqueous solutions \cite{isreal}.  In the case where the Debye length ($\lambda_d$) and separation ($z$) are small compared to the radius of the particle, the potential profile of the electrical double-layer interaction is well approximated by a decaying exponential\cite{de40}
\begin{align}
\label{eq:dl}
V_{dl} = A e^{-z/\lambda_d}
\end{align}
where $A$ is a constant which depends on the surface potentials and geometries of the interacting bodies \cite{p90,ve55} and can be quite large in the case of glass in water, and the Debye length, 
\begin{align}
\lambda_d = \sqrt{\frac{\epsilon_r \epsilon_0 kT}{e^2 N_A \sum_{i}c_i z_i}}
\end{align}
is determined by the permittivity of the fluid ($\epsilon_r$) and the concentration of electrolyte species in $mol/\textrm{m}^3$ ($c_i$) with valency $z_i$, where $N_A$ is the Avogadro constant.

As we lower the glass bead held in our optical tweezer in steps towards the glass surface, interaction of their double-layers can not only disturb the equilibrium position of the sphere relative to the optical tweezer focus, but can in some cases altogether prevent the sphere from coming into contact with the surface.

Figure \ref{fig:salt} illustrates this effect.  In the presence of a strong double layer, as the trapping beam focus moves towards the surface, the bead does not follow.  In the data, this manifests as lower than expected APD counts. However, by increasing electrolyte concentration, by the addition of NaCl to the colloidal mixture, the double layer is thinned and its influence diminished until the resulting intensity-height curve is similar to what one would expect from a hard-wall potential.  By examining the shape of the curves as a function of ion concentration, we determined 0.01M to be a reasonable lower bound on the required salt concentration for an accurate calibration.  As our piezo has precision better than 5 nm, we estimate that the contact-point determined in this way has an uncertainty less than 10 nm.

\section{Results}
\label{sec:results}

To demonstrate the utility of our calibration method we present two results, obtained with 3 $\mu$m diameter glass beads optically trapped in water above a glass surface.  The first (Figure \ref{fig:betas}) shows our method's sensitivity to changes in angle of incidence of the totally internally reflected (550 nm) beam.  The second (Figure \ref{fig:diff}) demonstrates simultaneous calibration as well as high spatial and temporal resolution particle tracking via a highly accurate, long range measurement of the hindered near-wall diffusion coefficient.

\subsubsection{Calibrated evanescent field profile}
\label{sec:betas}
\begin{figure}[htb]
\centering
\includegraphics[width=3in]{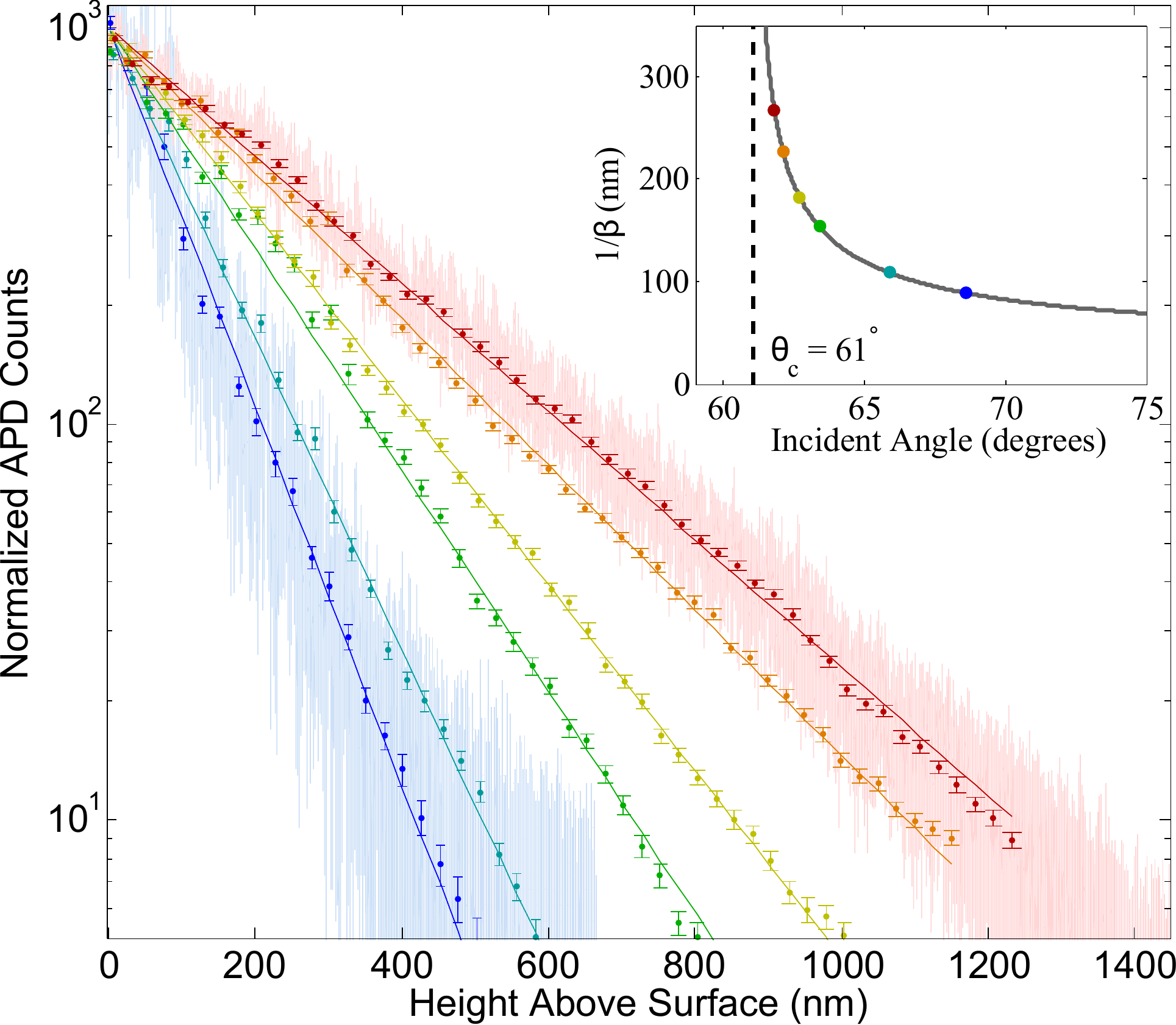}
\captionsetup{justification=raggedright,singlelinecheck=false}
\caption{ \footnotesize Normalized scattered intensity profiles as a function of height for a 3 $\mu$m diameter SiO$_2$ bead in water as angle of incidence of probe beam is varied.  Fitted decay lengths of the evanescent field intensity are 90.8, 110.6, 115.6, 159.9, 185.2, 236.4, 268.9 nanometers, corresponding to angles of 68.6, 65.9, 63.4 63.3, 62.7, 62.1, and 61.8 degrees respectively.  The error on all fits is less than 1\%.  The critical angle is 61.0 degrees for the system.  For clarity, raw data is only shown for the longest and shortest decay length measurements.  Graph inset plots color-coded measured decay lengths as dots on the curve $\beta^{-1}(\theta)$.
}%
\label{fig:betas}%
\end{figure}

The decay-length ($\beta^{-1}$) of the evanescent field used for positional sensing depends upon the angle of the incident beam and diverges as the angle approaches the critical angle ($\theta_c = \sin^{-1}(n_2/n_1)$) \cite{born99}. The functional form of $\beta$ versus angle of incidence is,
\begin{align}
\beta = \frac{4 \pi}{\lambda} \sqrt{(n_1 \sin\theta_1)^2 - n_2^2}
\end{align}

As it is difficult to conventionally determine the angle of incidence with precision better than half a degree, uncalibrated TIRM measurements, even in systems with well-behaved exponential intensity-distance relations, suffer from growing uncertainty in the $\beta$ parameter.  For example, assuming the typical quoted index of refraction for glass slides of 1.52 $\pm$ 0.01 and a half-degree uncertainty in angle, at $\beta^{-1}$ of 150 nm, the uncertainty in $\beta$ is approximately 13\%.  For a bead at a nominal height of 400 nm from the surface, this translates into an uncertainty of about 50 nm from the determination of $\beta$ alone.  As this error is systematic to the measurement, it cannot be reduced by increasing integration or data collection time.  At $\beta^{-1}$ of 250 nm, the uncertainty becomes even more dramatic ($\sim$35\%), corresponding to an uncertainty of 140 nm at a bead height of 400 nm.

As we mentioned in the introduction, one major advantage of the TIRM technique over other high-precision positional sensing techniques is the experimenter's ability to tune desired sensitivity against desired range by changing the $\beta^{-1}$ parameter.  This ability is lost in a typical uncalibrated TIRM experiment, where evanescent-field decay-lengths are usually kept to around 100 nm or below to ensure a correct estimation of the $\beta$ parameter.

Figure \ref{fig:betas} demonstrates precise and accessible tuning of the decay-length parameter to values between 90 and 270 nm, spanning the range from that of a typical TIRM experiment to the upper limit of the validity of the exponential-intensity profile \cite{he06}. The scans took approximately two minutes each and are fit to pure exponentials according to the least-squares method. The error on the $\beta$ parameters do not exceed 1\% on any fit and can be further improved with increased integration time.

\subsubsection{Wall Effects on Diffusion Coefficients}
\label{sec:diff}

A sphere in fluid diffusing close to a planar surface will experience increased viscous drag as compared to its motion in the bulk.  This increased drag force manifests in the Brownian dynamics of the particle as reduced diffusion.  In particular, the parallel $D_\parallel$ and perpendicular $D_\perp$ diffusion coefficients of a sphere of radius $R$ at height $z$ above a substrate calculated using hydrodynamic theory are\cite{b61,g67}
\begin{widetext}
\begin{eqnarray}
\label{eq:dpar}
D_{\parallel} &= &D_0 \left( 1- \frac{9}{16} \Lambda +\frac{1}{8} \Lambda^3 - \frac{45}{256} \Lambda^4 -\frac{1}{16}\Lambda^5\right)^{-1} \\
\label{eq:dperp}
D_{\perp} &= &D_0 \left[ \frac{4}{3} \sinh\alpha \sum_{n=1}^{\infty} \frac{n(n+1)}{(2n-1)(2n+3)}
\left( \frac{2\sinh(2n+1)\alpha + (2n+1)\sinh2\alpha}{4 \sinh^2(n+1/2)\alpha - (2n+1)^2 \sinh^2\alpha} -1 \right) \right]^{-1}
\end{eqnarray}
\end{widetext}
where $\Lambda = \frac{R}{R+z}$ and $\alpha = \text{cosh}^{-1}(\frac{z+R}{R})$.

The exact dependence of the diffusion coefficients on $z$ has been validated by experimenters with varying methods but overall success \cite{sc07,ca07}. As such measurements require very high temporal and positional resolution, for motion perpendicular to the planar surface, TIRM has more than once been the sensing technique of choice \cite{c01,be00,fr93,pa96,ki04}. In these measurements, however, TIRM has historically fallen short of the holographic, interferometric methods of particle tracking.  Without a reliable calibration scheme, errors in absolute position are large and the measurement range is limited.  Existing diffusion coefficient measurements based on TIRM do not extend beyond several hundred nanometers from the surface and are either sparse or systematically disagrees with hydrodynamic theory.

Here, we perform a simultaneous calibration and particle-tracking experiment where a 3 $\mu$m diameter glass sphere held by an optical tweezer is lowered in steps of 50 nm towards the glass surface in an evanescent field with a decay length of about 300 nm.  At each piezo position, the scattered field intensity is monitored for 20 s with a temporal resolution of 200 $\mu$s.  The average intensity at each point is used to build the intensity-position calibration curve, and the full time-series, converted to positions, is used to generate the autocorrelation function.

\begin{figure}[htb]
\centering
\includegraphics[width=3in]{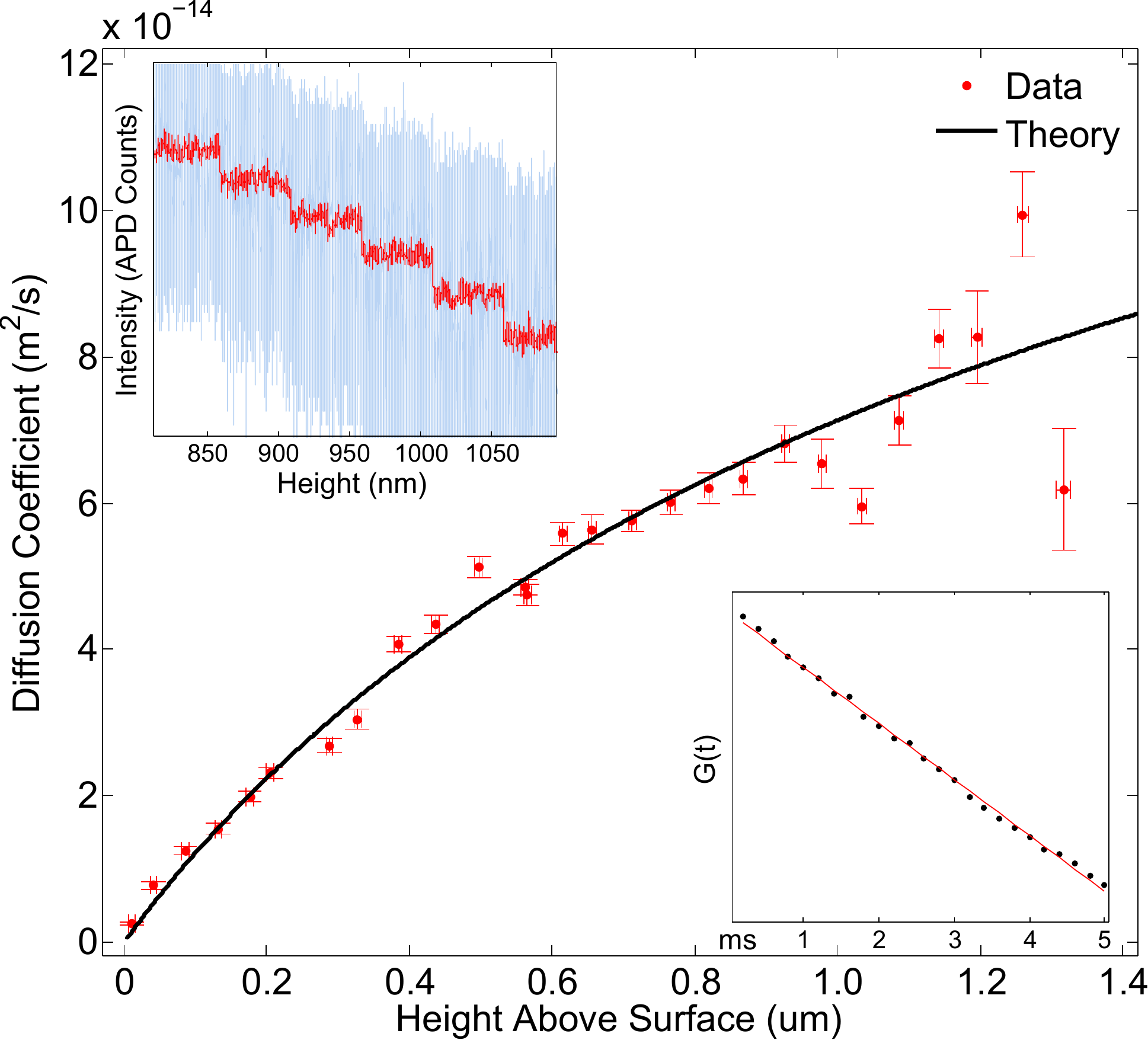}
\captionsetup{justification=raggedright,singlelinecheck=false}
\caption{ \footnotesize Measured diffusion coefficient as a function of height for motion in the normal direction.  Temperature assumed to be room temperature (22 C).  Fitted particle radius is (1.26 $\pm$ 0.02) $\mu$m.  \emph{Inset top left:}  Raw scattered intensity data (blue) and its moving average (red) showing 50 nm piezo steps.  \emph{Inset bottom right:} Example autocorrelation function (black dots) and linear fit (red line) in the intermediate time (diffusive) regime.
}%
\label{fig:diff}%
\end{figure}

The position autocorrelation function for a particle undergoing damped Brownian motion in a harmonic oscillator potential (formed by the optical trap) can be derived for instance via a simple Langevin equation\cite{uh30,wa45}.  For overdamped systems (such as ours) where the particle is weakly trapped in a viscous fluid, the autocorrelation function takes on a simple analytical form\cite{uh30} 
\begin{align}
\label{eq:auto}
G(t) &= <x(0) x(t)> \\ 
&= \frac{kT}{m w_0^2} e^{\frac{-\gamma t}{2}}\left(\text{cosh}(w_1 t) +\frac{\gamma}{2 w_1} \text{sinh}(w_1 t)\right) \nonumber  
\end{align}
where $t$ is the time delay between two measurements of particle position, $w_0$ is the resonant frequency of the trap, $w_1 = \sqrt{\gamma^2/4 - w_0^2}$, and $\gamma = 6 \pi R \eta / m$ is the Stokes drag coefficient divided by mass. In the limit of very small $t$, that is, at times shorter than the momentum relaxation time for the sphere (in our case around 1 $\mu$s), the motion of the particle is ballistic and the autocorrelation function is quadratic.  In the limit of large $t$, at $t$ $>$ 20 ms, the autocorrelation function reflects that of an overdamped harmonic oscillator and is a decaying exponential.   At intermediate time scales, the motion resembles that of a freely diffusing particle, and the auto-correlation can be approximated by a linear function,
\begin{align}
G(t) \approx - D_{\perp}t + \text{const}.
\end{align}
whose slope is the diffusion coefficient $D_\perp$. This dependence is shown in Figure \ref{fig:diff} (inset), which allows us to extract $D_\perp$ via a linear fit of the particle's autocorrelation function at $t$ between 200 $\mu$s and 5 ms.  For a more detailed description of the data analysis see the supplementary information. 

Figure \ref{fig:diff} shows the diffusion coefficient as a function of height obtained in this manner, which agrees well with predictions from hydrodynamic theory.  The measurement which produced this data took a total of 20 minutes and spans the distance range from a few nanometers to 1.2 microns above the glass surface.  

As the trapping power was measured to be 40 mW behind the objective, we assume negligible heating of the fluid and particle \cite{p03}.  The fitted radius of the bead assuming the measured room temperature of 22 C is 1.26 $\pm$ 0.02 $\mu$m.  

\subsubsection{Precise Measurement of Bead Radius}

To confirm the accuracy of the fit result, we implement  an independent method of determining bead radius, whereby we measure the change in effective bead mass as a function of trap radiation pressure and extract the bead mass intercept (at zero radiation power) via a linear regression.

The potential energy profile for a glass bead above a glass surface is expected to be, combining the double layer repulsion discussed previously and the gravitational and radiation pressure effects, \cite{p90,d91,f93}.  
\begin{align}
\label{eq:pot}
V(z)=A e^{-z/\lambda_d} + Bz
\end{align}
where $B = (\rho_{\text{bead}}-\rho_{\text{water}})V g + \phi$, $\rho_{\text{bead}}$ and $\rho_{\text{water}}$ are the densities of glass and water, $V$ the volume of the bead, $g$ the gravitational acceleration, and $\phi$ the force due to radiation pressure.  We reduce the NA of the trapping beam by closing the iris in the back focal plane of the objective, largely removing axial confinement of the particle, and obtain the parameter B by fitting the potential for several different trapping powers.  Assuming that radiation pressure scales linearly with laser power (P) for a given bead and trap configuration, i.e. $\phi = \alpha P$, a linear fit of B vs. P yields from the intercept the bead mass, and therefore radius.

\begin{figure}[htb]
\centering
\includegraphics[width=3in]{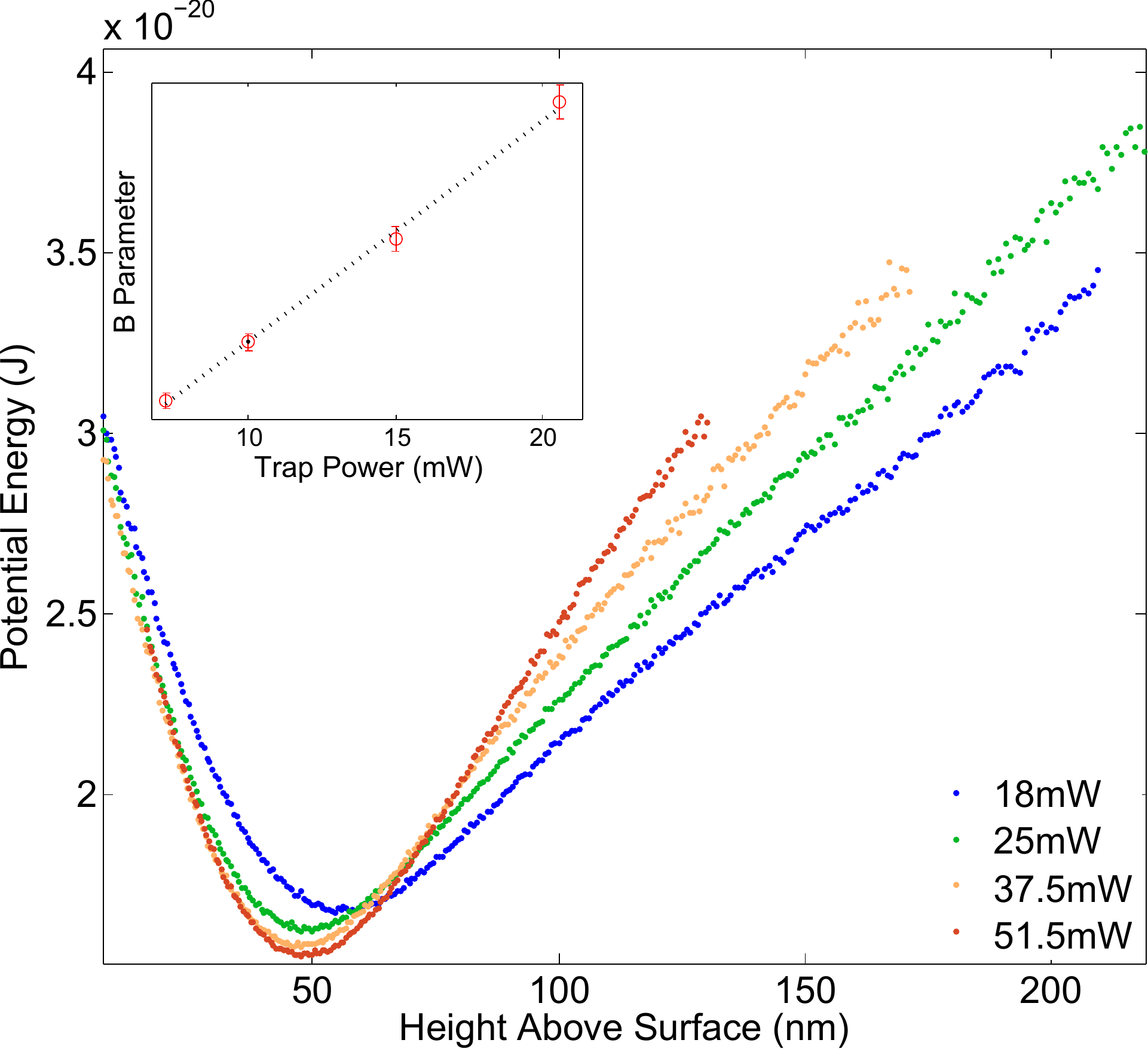}
\captionsetup{justification=raggedright,singlelinecheck=false}
\caption{ \footnotesize Potential energy profiles measured for 4 different powers of the low numerical aperture trapping beam. Curves are shifted horizontally to better visualize change in the B parameter (as defined in Equation \ref{eq:pot}), or the slope of the linear portion of the potential.  Inset: B parameter vs. trapping power, and the linear fit used to calculate the particle's radius.
}%
\label{fig:pot}%
\end{figure}

Figure \ref{fig:pot} shows potential energy profiles obtained for the glass sphere whose diffusion coefficient we measured in 
Section \ref{sec:diff}.   The optical trap NA was reduced to less than 0.2 and the trap power varied between 18 mW and 51 mW.  The bead radius we obtain by fitting the potential energy profiles, assuming room temperature, is 1.27 $\pm$ 0.05 $\mu$m, which is in good agreement with the 1.26 $\pm$ 0.02 $\mu$m result from \ref{sec:diff}.

\section{Conclusions}

We have developed a method for high resolution absolute position particle-tracking using the TIRM sensing method.  As the TIRM technique is itself highly sensitive and highly tunable, our main innovation is the introduction of an in-situ optical tweezer calibration that harnesses the full potential of the measurement technique.  By direct measurement of the essential experimental parameters under exact experimental conditions, we eliminate errors associated with their usual estimation or calculation.  

We demonstrate a more than 10x error reduction in determining the decay-length parameter which sets the measurement range of a TIRM experiment.  We show free and ready tuning of this parameter, a valuable experimental degree of freedom rarely before exploited due to prohibitively large errors.  We are able to locate the point of contact, with better than 10 nm precision, yielding absolute instead of relative positional values.  To show the extended versatility of the TIRM technique we repeated an experiment requiring high sensitivity, long range, and absolute positions that has previously proved challenging for this sensing method.  Our results exceed in range that of previous experiments by a factor of 3 and show good agreement with hydrodynamic theory throughout.  In the future, this calibration technique can be generalized to ultra-long range or systems involving metallic surfaces which would allow for the precise measurements of novel optical forces in previously unexplored systems.

\section{Acknowledgements}
This work is supported by NSF GFRP grant number DGE1144152.  We kindly thank the Evelyn Hu and David Weitz groups at Harvard for shared equipment and helpful discussions. 

\bibliographystyle{unsrt}

\bibliography{method_paper}

\end{document}